# STRUCTURAL CHARACTERIZATION OF MAGNETIC NANO-PARTICLES SUSPENSIONS, USING MAGNETIC MEASUREMENTS

*Nicolae Calin POPA [1], Ali SIBLINI [2], Jean Jacques ROUSSEAU [2],*
*Marie-Françoise BLANC-MIGNON [2], and, Jean Pierre CHATELON [2]*

[1] Center for Advanced and Fundamental Technical Research (CAFTR), Romanian Academy –
Timisoara Branch, Bd. Mihai Viteazul nr 24, 300223 Timisoara, Romania,
Phone: + 4 0256 403703, Fax: + 4 0256 403700,
E-mail: ncpopa@acad-tim.tm.edu.ro, or, ncpopa@flumag2.mec.utt.ro.

[2] DIOM, Dispositifs et Instrumentation en Optoélectronique et Micro-ondes, Université Jean Monnet,
23 rue du Dr. P. Michelon, 42023 Saint - Etienne, Cedex 2, France,
Phone (33) 0477 485079, Fax: (33) 0477 485039,
E-mail: Ali.Siblini@univ-st-etienne.fr, or, rousseau@univ-st-etienne.fr.

**ABSTRACT**

The paper describes some characteristics of the "P" curves for structural characterization of magnetic nano-particles suspensions (complex fluids, complex powders, complex composite materials, or living biological materials having magnetic properties). In the case of these materials, the magnetic properties are conferred to various carrier liquids by artificially integrating in their structure ferromagnetic particles of different sizes. The magnetic properties are usually shown by the hysteresis curve. The structure can be seen on (electronic) micrography. The P curves offer another possibility to determine the structure of the magnetic component of a complex fluid by numerical analysis of the magnetization curve experimentally obtained. The paper presents a detailed approach of the P curves and some limitations in their use.

## 1. INTRODUCTION

Usually, magnetizable fluids include nano-magnetic fluids (called also magnetic fluids or ferrofluids, [1 - 11]), magneto – rheological fluids ([12 - 14]), or other fluids obtained mixing the first two categories. In all these complex fluids (suspensions), their magnetic properties are obtained by artificially integrating, in the mass of different carrier liquids, ferromagnetic particles (of different sizes and/or materials). The magnetic properties of these complex fluids are shown by the hysteresis curve, and, their structure can be seen on the (electronic) micrography. In a previous paper ([15]), we defined the "P" (peak) curves, which can deliver information concerning the structure of magnetizable fluids, by numerical derivation (relative to the magnetic field strength H) of the magnetization curves relative to the saturation magnetizations:

$$P(H) = \frac{d[M(H)/M_S]}{dH}, \quad (1)$$

where $P$ is the P value, $H$ is the magnetic field strength, $M$ is the fluid magnetization, and, the subscript S denotes the saturation point from the magnetization curve. We defined also the "P shape factor" (PSF) for the P curves in solid matrix:

$$PSF = \frac{P(0)_{par}}{P(0)_{perp}}, \quad (2)$$

where the subscript *par* or *perp* denotes that we determined the P curves with the measurement of magnetic field respectively parallel or perpendicular to the solidification magnetic field.

We showed ([15]) that, for the same type of magnetic nano-particles (the same magnetic material and the same technology to obtain the particles), the height of the P curves is greater when the magnetic particles are smaller, and, the spread of the P curves is wider when the magnetic particles are of greater dimensions. We showed also that the P curves are the same if we use the volume magnetization or the mass magnetization; the P curves do not depend on the accuracy of the magnetometer y-axis calibration, and on the accuracy of sample mass or volume measurement. From all these, we get the most interesting application for P curves, the possibility to investigate the living biological material (from magnetic point of view) without extracting a sample.

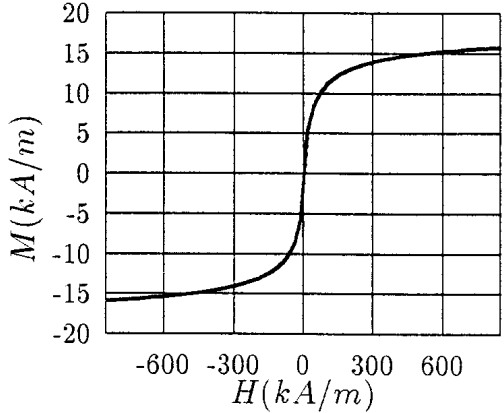

Fig. 1: Typical hysteresis curve for a nano-magnetic fluid (experimentally measured).

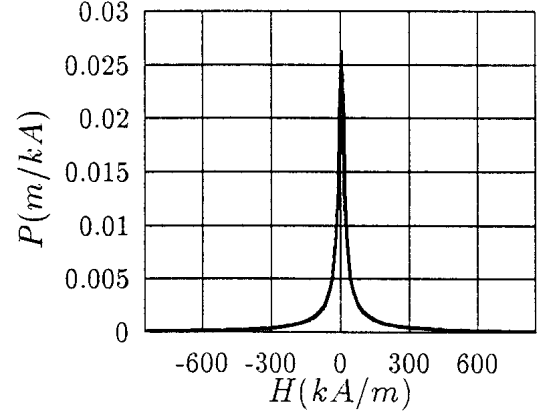

Fig. 2: P curve for the nano-magnetic fluid from Fig. 1.

Fig. 1 presents a typical hysteresis curve (experimentally obtained using a Vibrating Sample Magnetometer VSM 880, ADE Technologies, USA) for a nano-magnetic fluid. The fluid (Ferrofluidics, USA) has the saturation magnetization $M_S$ = 15.9 kA/m, and the particles diameter 8.5 nm (ferrite). As it is known, the nano-magnetic fluids have not hysteresis loop. Fig. 2 shows the P curve for the nano-magnetic fluid from Fig. 1.

The paper presents a detailed approach of the P curves. We tried to find mathematical equations to approximate the P curves, and, to establish some limitations in the use of these P curves.

## 2. DETAILED APPROACH OF "P" CURVES

To better understand the theoretical and practical possibilities offered by the P curves, some of their characteristics will be presented.

I. From the hysteresis curve of magnetizable fluids, where $M(H) = -M(-H)$ ($M \geq 0, H \geq 0$), we can see that the P curves are symmetrically versus the Oy axis, that meaning $P(H) = P(-H)$ ($P \geq 0, H \geq 0$) (Figs. 2 and 7). We can observe that,

$$\int_{-\infty}^{+\infty} P(H) \cdot dH = 2 \cdot \int_{0}^{H_S} P(H) \cdot dH . \qquad (3)$$

II. For an arbitrary point denoted by the subscript X,

$$\int_{0}^{H_X} P(H) \cdot dH = \frac{M_X}{M_S} . \qquad (4)$$

III. If the arbitrary point is the saturation point,

$$\int_{0}^{H_S} P(H) \cdot dH = 1 . \qquad (5)$$

This means that for any P curve, the area between the P curve, and, the Oy and Ox axis is always 1. With other words, the amplitude and the spread of a P curve are not independent. In this situation, the question is if a P curve can be theoretically defined through a mathematical equation which use several parameters, or through a mathematical relation having one parameter only (this parameter being P(0)). The P curve defines, in some limits, the nano-magnetic particles. If we can express the P curve using one parameter only, than, having a P curve (experimentally obtained), from this P curve we can obtain information about only one parameter of the nano-magnetic particles. This parameter can be a physical parameter (e.g. the most frequent dimension of the particles), or can be a mathematical relation (e.g. a ratio, a sum, etc.) between other physical characteristics of the nano-particles.

IV. Because the P values are the result of numerical derivation of some experimental measurements, the calculus of these values is performed with certain approximation. In an arbitrary point, we can calculate the derivative to the left, to the right, or their arithmetical mean (center derivative). Fig. 3 presents the difference between these three situations (P values for a ferrofluid, [3], having $M_S$ = 15.6 kA/m, and, the particles diameter of 10.5 nm). For a general view of P curves (Figs. 2 and 7), usually we used the center derivative. Further, for a theoretical approximation of the P curves, it is preferable to use the derivative to the right.

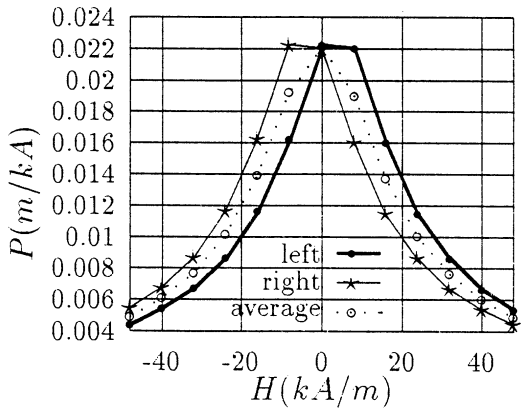

Fig. 3: Three different possibilities to calculate the derivative in an arbitrary point: derivative to the left, to the right, and, the arithmetical mean of the first two (center derivative).

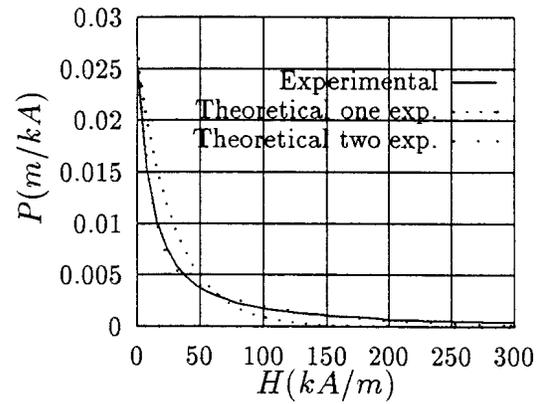

Fig. 4: P curve (for the nano-magnetic fluid from Fig. 1) and its theoretical approximations.

V. At the first sight, we have a great temptation to approximate the P curve, using the theoretical relation:

$$P_t(H) = P(0) \cdot e^{-\frac{H}{H^*}}. \qquad (6)$$

were, the subscript "t" denotes a theoretical curve.

If we admit this approximation (for $0 \leq H \leq H_S$), than

$$\int_0^{H_X} P_t(H) \cdot dH = -H^* \cdot [P_t(H_X) - P(0)]. \qquad (7)$$

In fact, at the saturation point $P(H_S) = 0$, so we can write

$$\int_0^{H_S} P_t(H) \cdot dH = H^* \cdot P(0). \qquad (8)$$

Taking into consideration the equalities (8) and (5),

$$H^* = \frac{1}{P(0)}, \qquad (9)$$

and, the approximation (6) becomes

$$P_t(H) = P(0) \cdot e^{-P(0) \cdot H}. \qquad (10)$$

This means that, if we admit the approximation (6), the P curves are completely determined by the value *P(0)*. In this situation, the P curves can deliver information about only one parameter of the magnetizable nano-fluid (this parameter could be the most frequent dimension of the particles).

## 3. THEORETICAL APPROXIMATIONS OF "P" CURVES

After many experimental measurements and theoretical approximations, we concluded that, the approximation given by relation (6) (or its particular case given by equation (10)) is not good enough for the P curves. Fig. 4 shows a P curve and its theoretical approximation obtained using one exponential function only (relation (6) for H* = 30).

I. The same figure presents the approximation (of the P curve) obtained using a sum of two exponential functions, *E₁(H)* and *E₂(H)*.

$$E_1(H) = E_1(0) \cdot e^{-\frac{H}{H_1^*}}, \qquad (11)$$

$$E_2(H) = E_2(0) \cdot e^{-\frac{H}{H_2^*}}, \qquad (12)$$

$$P_t(H) = E_1(H) + E_2(H). \qquad (13)$$

In this case, the theoretical curve (for: *E₁(0) = 0.75P(0)*, *H*₁ = 10 kA/m*, *E₂(0) = 0.25P(0)*, and, *H*₂ = 85 kA/m*) is almost superposed to the experimental curve. We determined these four parameters for the equation (13) using numerical investigations.

II. We get a very good theoretical (analytical) approximation of P curves using exponential functions defined for the interval between two successive experimental points. So, we take into consideration the values $H_o$ $(=0)$, $H_1$, $H_2$, ..., $H_n$, and, $H_{n+1}$, of magnetic field, where, for the P curve, we experimentally obtained respectively the values $v_o$ $(=P(0))$, $v_1$, $v_2$, ..., $v_n$ (for $i = 1$ to $n$, all $v_i > 0$), and, $v_{n+1}=0$.

We define the function $F_i(H)$,

$$F_i(H) = \begin{cases} 0 & \text{for } H < H_i, H_{i+1} \leq H \\ v_i \cdot e^{-\frac{H-H_i}{H_i^*}} & \text{for } H_i \leq H < H_{i+1} \end{cases}$$

where: $H_i^* = \dfrac{H_{i+1} - H_i}{\ln v_i - \ln v_{i+1}}$

(14)

The expression for $H_i^*$ results from the condition (15).

$$\lim_{\substack{H \to H_{i+1} \\ H < H_{i+1}}} F_i(H) = v_i \cdot e^{-\frac{H_{i+1}-H_i}{H_i^*}} = v_{i+1}. \quad (15)$$

(This means, the sum of n consecutive functions, gives a continuous function.) For the continuity, we remark that

$$F_i(H_i) = v_i. \quad (16)$$

We define also the linear interpolation function $G_i(H)$,

$$G_i(H) = \begin{cases} 0, & \text{for, } H < H_i, H_{i+1} \leq H \\ v_i + \dfrac{\Delta v_i (H - H_i)}{\Delta H_i}, & \text{for, } H_i \leq H < H_{i+1} \end{cases}$$

where: $\Delta v_i = v_{i+1} - v_i$, and, $\Delta H_i = H_{i+1} - H_i$

(17)

With these from above, the approximation of P curve is

$$P_t(H) = \sum_{i=0}^{n-1} F_i(H) + G_n(H). \quad (18)$$

III. We get also an acceptable approximation using the relation

$$P_t(H) = \sum_{i=0}^{n} G_i(H). \quad (19)$$

IV. Another way to obtain a good analytical approximation of P curve is to approximate the P curve using the sum of two exponential functions, this sum passing through four experimental points (relations (11), (12), and (13), but, the four parameters will be analytically determined). For that we use four equations having the form

$$E_1(H_i) + E_2(H_i) = v_i, \quad (20)$$

where, $i = 1$ to $4$, $H_0 = 0$, and, $v_i = P(H_i)$.

The obtained equations system is

$$\begin{cases} E_1(0) + E_2(0) = v_0 \\ E_1(0) \cdot e^{-\frac{H_1}{H_1^*}} + E_2(0) \cdot e^{-\frac{H_1}{H_2^*}} = v_1 \\ E_1(0) \cdot e^{-\frac{H_2}{H_1^*}} + E_2(0) \cdot e^{-\frac{H_2}{H_2^*}} = v_2 \\ E_1(0) \cdot e^{-\frac{H_3}{H_1^*}} + E_2(0) \cdot e^{-\frac{H_3}{H_2^*}} = v_3 \end{cases} \quad (21)$$

To solve the system, we note

$$e^{-\frac{H_1}{H_1^*}} = x, \quad \text{and,} \quad e^{-\frac{H_1}{H_2^*}} = y. \quad (22)$$

It doesn't mater which four experimental points we consider to belong to the sum of the two exponential functions. So, we consider an arbitrary $H_1$, and,

$$H_2 = 2 \cdot H_1, \text{ and,} \quad H_3 = 3 \cdot H_1. \quad (23)$$

We get the system

$$\begin{cases} E_1(0) + E_2(0) = v_0 \\ E_1(0) \cdot x + E_2(0) \cdot y = v_1 \\ E_1(0) \cdot x^2 + E_2(0) \cdot y^2 = v_2 \\ E_1(0) \cdot x^3 + E_2(0) \cdot y^3 = v_3 \end{cases} \quad (24)$$

From the first equation of the system (24)

$$E_2(0) = v_0 - E_1(0). \quad (25)$$

We replace the value of $E_2(0)$ in the other equations, and, from the second equation

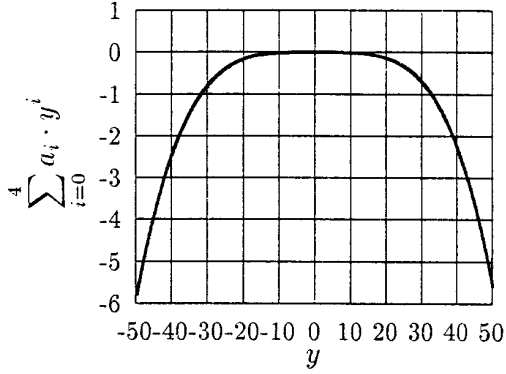

Fig. 5: General aspect of the function described by equation (28).

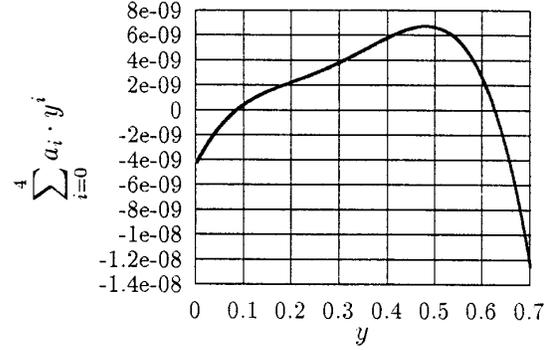

Fig. 6: Detail of Fig. 5.

$$E_1(0) = \frac{v_1 - v_0 \cdot y}{x - y}. \quad (26)$$

On the same way, from the third equation we can still obtain a unique analytical expression for x,

$$x = \frac{v_2 - v_1 \cdot y}{v_1 - v_0 \cdot y}. \quad (27)$$

The last equation of the system (24) becomes an equation where $y$ has exponent up to 4,

$$\begin{aligned}
\sum_{i=0}^{4} a_i \cdot y^i &= 0 \\
a_4 &= v_0\left(v_1^2 - v_0 v_2\right) \\
a_3 &= v_0 v_1 v_2 - 2v_1^3 + v_0^2 v_3 \\
a_2 &= 3v_1\left(v_1 v_2 - v_0 v_3\right) \\
a_1 &= 2v_1^2 v_3 - 3v_1 v_2^2 + v_0 v_2 v_3 \\
a_0 &= v_2\left(v_2^2 - v_1 v_3\right)
\end{aligned} \quad (28)$$

The problem is that, even the equation (28) exists for any four points from the plane (the first point having abscissa 0, the second having an arbitrary abscissa, and, the next two abscissas respecting the relation (23)), not through any four points from the plane we can pass the sum of the same two exponential functions. Therefore, the equation (28) has real solutions for some "four points" ($v_i$), which must respect some conditions. It is very difficult to analytically express these conditions. It is easier to find numerically the real solutions of the equation (28). After many experiments, we concluded that if the four points belong to a P curve, the equation (28) has real solutions. We replace these solutions of equation (28) in (27), and than consecutively we solve (26) and (25). From (22) we get

$$H_1^* = -\frac{H_1}{\ln x}, \quad \text{and,} \quad H_2^* = -\frac{H_1}{\ln y}. \quad (29)$$

Fig. 5 presents a general aspect of the function described by equation (28), and, Fig. 6 shows a detailed aspect (of the same function) where we can see the real solutions of this equation. For these two figures, we considered four points of the P curve from Fig. 1. ($H_0 = 0$ kA/m, $H_1 = 40$ kA/m, $H_2 = 80$ kA/m, $H_3 = 120$ kA/m, $v_0 = 2.627385 \times 10^{-2}$ m/kA, $v_1 = 5.278942 \times 10^{-3}$ m/kA, $v_2 = 2.379331 \times 10^{-3}$ m/kA, and, $v_3 = 1.412452 \times 10^{-3}$ m/kA.)

## 4. SOME LIMITATIONS IN THE USE OF "P" CURVES

By experimental measurements, we concluded that the applicability superior limit for P curves in magnetic suspensions is for the particles having diameters of about 20 - 25 μm. Over this limit, the P curves (of the complex fluids having different particles dimension) have the same aspect. The applicability inferior limit is at the dimension where the particles lose their magnetic properties.

All the magnetizable fluids presented above, for the magnetic particles, have a dimensional distribution curve with only one maximum. Using two different complex fluids of this type (further on denoted as the first and the second fluid), we prepared a new magnetizable fluid having two maximums in its dimensional distribution curve.

The first fluid had the particles diameter of about 10.5 nm (nano-magnetic fluid with ferrite particles), $M_S = 15.6$ kA/m ([3]), and, $P(0) = 0.022$. The second fluid had the particles diameter of about 25 μm (Fe particles), $M_S = 723$ kA/m (Hoeganaes Europe, Romania), and, $P(0) = 0.004$.

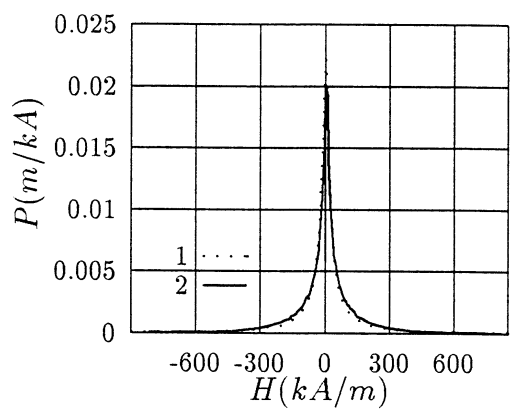

Fig. 7: P curves for the first fluid (denoted on the figure by 1 and having only one maximum in the dimensional distribution curve), and, for the new fluid (denoted on the figure by 2 and having two maximums in the dimensional distribution curve).

For the new fluid, we used about 97 % from the first fluid, and, 3 % from the second fluid. The new fluid has $M_S$ = 50.3 kA/m and $P(0)$ = 0.02.

Fig. 7 presents the P curves for the first and for the new fluid. We can observe that the two P curves from Fig. 7 are very close. From the P curve of the new fluid, we can't conclude that it belongs to a fluid having two maximums in the dimensional distribution curve.

We can conclude that, the application of the P curves is recommended at the magnetizable complex fluids (powders, suspensions, etc.) having only one maximum in the dimensional distribution curve of magnetic particles.

## 5. ACKNOWLEDGEMENTS